\documentstyle[aps]{revtex}

\begin{document}
\title{Preparing remotely two instances of quantum state}
\author{Ya-fei Yu$^{a\thanks{%
Corresponding author. E-mail address: yfyu@wipm.ac.cn}}$, Jian Feng$^{b}$
and Ming-sheng Zhan$^{a}$}
\address{$^{a}$State Key Laboratory of Magnetic Resonance and Atomic and Molecular\\
Physics,Wuhan Institute of Physics and Mathematics, Chinese Academy of\\
Sciences,\\
Wuhan 430071, PR China\\
$^{b}$Institute of Optical Communication, Liaocheng Teachers University,\\
Liaocheng 252059, Shandong, PR China }
\maketitle

\begin{abstract}
In this short note, we propose a scheme, in which two instances of an
equatorial state (or a polar state) can be remotely prepared in one-shot
operation to different receivers with prior entanglement and 1 bit of
broadcasting. The trade-off curve between the amount of entanglement and the
achievable fidelity is derived.

Key words: remote state preparation, entanglement

PACS number(s): 03.67.HK
\end{abstract}

\section{Introduction}

Quantum entanglement and classical communication are two elementary
resources in quantum information field. The trade-off relation between
entanglement and classical communication is inspected in many quantum
information processing. For example, in quantum teleportation \cite
{teleportation} the utilization of quantum entanglement reduces greatly the
cost of classical communication. Compared with quantum teleportation, remote
state preparation (RSP)\cite{pati,lo,bennett,igor,zheng,shi} exhibits the
stronger trade-off between the required entanglement and the classical
communication cost \cite{bennett,igor}. It is found that in the presence of
a large amount of prior entanglement, the asymptotic classical communication
cost of RSP for general states is 1 bit per qubit, half that of
teleportation. On the other hand, the prior knowledge about quantum state in
the process of quantum information reduces the communication cost including
entanglement and classical communication. For instance in quantum remote
control \cite{pra63}, for saving the communication requirement the
teleportation of angles is divided into two cases of rotations by $\pi $
around any axis lying within the equatorial plane and arbitrary rotations
around the $z$ axis \cite{pra65}. As a special case of quantum remote
control, the RSP strongly displays the trade-off relation between the
restriction and the communication requirement: if only 1 ebit of
entanglement is supplied, for general states the classical communication
cost in RSP is equivalent to that in standard teleportation \cite{lo};
however, for constrained ensemble of states, the one is less than that in
quantum teleportation\cite{pati,lo}.

Here we devise a scheme to generalize the RSP to one receiver to the case of
multiple receivers. And differently from the previous work \cite
{pati,lo,bennett,igor,zheng,shi}, we investigat the connection between the
entanglement and the achievable fidelity under the conditions of certain
cost of classical communication and constrained state. For simplicity, we
consider the case of two receivers, in which Alice, using prior entanglement
pre-shared with Bob and Charlie and broadcasting 1bit of classical
information to Bob and Charlie, helps Bob and Charlie to recreate
simultaneously an equatorial state (or a polar state) in their respective
lab. We derive the trade-off curve between the amount of entanglement and
the achievable fidelity. The detail is elaborated in Section II, and the
results are summarized in Section III.

\section{Preparing two instances of quantum state remotely}

The problem is that Alice, using prior entanglement shared with Bob and
Charlie and broadcasting 1 bit classical information to Bob and Charlie,
aims to prepare remotely in one-shot operation an instance of the equatorial
state to Bob and Charlie in their respective lab. The state to be remotely
prepared has the form of $\left| \varphi \right\rangle =\frac{1}{\sqrt{2}}%
(\left| 0\right\rangle +e^{i\phi }\left| 1\right\rangle )$, where $\phi \in
(0,2\pi ]$ is known to Alice, but unknown to Bob and Charlie.

According to the quantum cloning for equatorial qubit \cite{heng}, a
deliberate entangled state 
\begin{equation}
\left| \psi \right\rangle =\frac{1}{\sqrt{2}}(\left| 0\right\rangle
_{a}\left| \phi _{1}\right\rangle _{ABC}-\left| 1\right\rangle _{a}\left|
\phi _{0}\right\rangle _{ABC})
\end{equation}
is distributed among Alice, Bob and Charlie to prepare remotely two
instances of the equatorial state $\left| \varphi \right\rangle $. The
tripartite states $\left| \phi _{0}\right\rangle $ and $\left| \phi
_{1}\right\rangle $ are defined as 
\begin{eqnarray}
\left| \phi _{0}\right\rangle _{ABC} &=&\frac{1}{\sqrt{2}}\left|
0\right\rangle _{A}\left| 0\right\rangle _{B}\left| 0\right\rangle _{C}+%
\frac{1}{2}\left| 1\right\rangle _{A}\left| 0\right\rangle _{B}\left|
1\right\rangle _{C}+\frac{1}{2}\left| 1\right\rangle _{A}\left|
1\right\rangle _{B}\left| 0\right\rangle _{C}, \\
\left| \phi _{1}\right\rangle _{ABC} &=&\frac{1}{\sqrt{2}}\left|
1\right\rangle _{A}\left| 1\right\rangle _{B}\left| 1\right\rangle _{C}+%
\frac{1}{2}\left| 0\right\rangle _{A}\left| 0\right\rangle _{B}\left|
1\right\rangle _{C}+\frac{1}{2}\left| 0\right\rangle _{A}\left|
1\right\rangle _{B}\left| 0\right\rangle _{C}.  \nonumber
\end{eqnarray}
Alice is in possession of the qubits A and $a$, the qubits B and C belong to
Bob and Charlie, respectively. For RSP, Alice measures the qubit $a$ in the
basis $\left\{ \left| \varphi \right\rangle _{a},\left| \varphi ^{\bot
}\right\rangle _{a}\right\} $ where $\left| \varphi ^{\bot }\right\rangle
_{a}$ denotes the state orthogonal to $\left| \varphi \right\rangle _{a}$.
The basis $\left\{ \left| \varphi \right\rangle _{a},\left| \varphi ^{\bot
}\right\rangle _{a}\right\} $ is related to the old basis $\left\{ \left|
0\right\rangle _{a},\left| 1\right\rangle _{a}\right\} $ in the following
manner 
\begin{eqnarray}
\left| 0\right\rangle _{a} &=&\frac{1}{\sqrt{2}}(\left| \varphi
\right\rangle _{a}+e^{i\phi }\left| \varphi ^{\bot }\right\rangle _{a}), \\
\left| 1\right\rangle _{a} &=&\frac{1}{\sqrt{2}}(e^{-i\phi }\left| \varphi
\right\rangle _{a}-\left| \varphi ^{\bot }\right\rangle _{a}).  \nonumber
\end{eqnarray}
Rewriting the entangled state $\left| \psi \right\rangle $ in the basis $%
\left\{ \left| \varphi \right\rangle _{a},\left| \varphi ^{\bot
}\right\rangle _{a}\right\} $ gives 
\begin{eqnarray}
\left| \psi \right\rangle &=&\frac{1}{\sqrt{2}}(\left| 0\right\rangle
_{a}\left| \phi _{1}\right\rangle _{ABC}-\left| 1\right\rangle _{a}\left|
\phi _{0}\right\rangle _{ABC})  \nonumber \\
&=&\frac{1}{\sqrt{2}}(\frac{1}{\sqrt{2}}(\left| \varphi \right\rangle
_{a}+e^{i\phi }\left| \varphi ^{\bot }\right\rangle _{a})\left| \phi
_{1}\right\rangle _{ABC}-\frac{1}{\sqrt{2}}(e^{-i\phi }\left| \varphi
\right\rangle _{a}-\left| \varphi ^{\bot }\right\rangle _{a})\left| \phi
_{0}\right\rangle _{ABC})  \nonumber \\
&=&\frac{1}{\sqrt{2}}(\left| \varphi ^{\bot }\right\rangle _{a}\otimes \frac{%
1}{\sqrt{2}}(\left| \phi _{0}\right\rangle _{ABC}+e^{i\phi }\left| \phi
_{1}\right\rangle _{ABC})-e^{-i\phi }\left| \varphi \right\rangle
_{a}\otimes \frac{1}{\sqrt{2}}(\left| \phi _{0}\right\rangle _{ABC}-e^{i\phi
}\left| \phi _{1}\right\rangle _{ABC})).
\end{eqnarray}

If the result of the measurement gives $\left| \varphi ^{\bot }\right\rangle
_{a}$, three parties Alice, Bob and Charlie share a tripartite state $\left|
\xi \right\rangle _{ABC}=\frac{1}{\sqrt{2}}(\left| \phi _{0}\right\rangle
_{ABC}+e^{i\phi }\left| \phi _{1}\right\rangle _{ABC})$ as expected. Thus
the reduced density matrices on the qubits B and C are given as 
\begin{eqnarray}
\rho _{B} &=&Tr_{A,C}(\left| \xi \right\rangle _{ABC}\left\langle \xi
\right| )=\rho _{C}=Tr_{A,B}(\left| \xi \right\rangle _{ABC}\left\langle \xi
\right| ) \\
&=&\frac{1}{2}(\left| 0\right\rangle \left\langle 0\right| +\left|
1\right\rangle \left\langle 1\right| )+\frac{1}{\sqrt{2}}(e^{-i\phi }\left|
0\right\rangle \left\langle 1\right| +e^{i\phi }\left| 1\right\rangle
\left\langle 0\right| )  \nonumber \\
&=&(\frac{1}{2}+\frac{1}{2\sqrt{2}})(\left| \varphi \right\rangle
\left\langle \varphi \right| )+(\frac{1}{2}-\frac{1}{2\sqrt{2}})(\left|
\varphi ^{\perp }\right\rangle \left\langle \varphi ^{\perp }\right| ). 
\nonumber
\end{eqnarray}
In the view of the RSP of single-qubit state, this means that Alice assists
Bob and Charlie in preparing an approximation to the equatorial qubit state $%
\left| \varphi \right\rangle $, simultaneously and respectively. The
fidelity of the approximation with respect to the equatorial state is $\frac{%
1}{2}+\frac{1}{2\sqrt{2}}$, which can be proved to be optimal in this scheme
of preparing remotely two instances of the equatorial state. And the
fidelity of $\frac{1}{2}+\frac{1}{2\sqrt{2}}$ is also the optimal one of $%
1\rightarrow 2$ cloning of the equatorial state \cite{heng}.

Otherwise, if the result of the single-particle measurement is $\left|
\varphi \right\rangle _{a}$, we get an incorrect tripartite state, from
which the correct state $\left| \xi \right\rangle _{ABC}$ can be recovered
by performing Pauli operator $\sigma _{z}$ on each of the qubits A, B and C.

On the completion of the single-particle measurement, Alice broadcasts her
result of the measurement through public channel to Bob and Charlie. Then
depending on the result, the parties determine to rotate respective qubits
or do nothing. Consequently, the scheme can be applied to prepare remotely
and simultaneously two instances of an equatorial state to different
receivers.

It is observed that in the cut of $a:B$ (or $a:C$) of the entangled state $%
\left| \psi \right\rangle $, the relative entropy of entanglement \cite
{pra57,prl78}$E_{r}=\min_{\sigma \in D}S(\rho //\sigma )\approx 0.6095$ $%
ebit<1$ $ebit$, where $S(\rho //\sigma )=Tr\{\rho (\log \rho -\log \sigma
)\} $ is the quantum relative entropy, and the minimum is taken over D, the
set of separable states. So the fidelity obtained in our scheme is less than
1 because of the limit of the lower bound to the necessary resources in the
perfect RSP.

The scheme also succeeds in preparing remotely a polar state with the form
of $\left| \varphi ^{/}\right\rangle =\cos (\theta )\left| 0\right\rangle
+\sin (\theta )\left| 1\right\rangle ,$ where $\theta \in (0,\pi ]$ is known
only to Alice. Now the deliberate entangled state shared among Alice, Bob
and Charlie is previously proposed in \cite{teleclone}. That is, the states $%
\left| \phi _{0}\right\rangle $ and $\left| \phi _{1}\right\rangle $ in Eq.
(1) are defined as 
\begin{equation}
\left| \Phi _{0}\right\rangle _{ABC}=\sqrt{\frac{2}{3}}\left| 0\right\rangle
_{A}\left| 0\right\rangle _{B}\left| 0\right\rangle _{C}+\sqrt{\frac{1}{6}}%
\left| 1\right\rangle _{A}(\left| 0\right\rangle _{B}\left| 1\right\rangle
_{C}+\left| 1\right\rangle _{B}\left| 0\right\rangle _{C})
\end{equation}
\begin{equation}
\left| \Phi _{1}\right\rangle _{ABC}=\sqrt{\frac{2}{3}}\left| 1\right\rangle
_{A}\left| 1\right\rangle _{B}\left| 1\right\rangle _{C}+\sqrt{\frac{1}{6}}%
\left| 0\right\rangle _{A}(\left| 0\right\rangle _{B}\left| 1\right\rangle
_{C}+\left| 1\right\rangle _{B}\left| 0\right\rangle _{C}).
\end{equation}
Similarly, when the single-particle measurement gives $\left| \varphi ^{\bot
}\right\rangle _{a}$, three parties do nothing to their qubit; or they apply
a rotation operator $-i\sigma _{y}$ on their respective qubit to retrieve
the desired tripartite state. Resultantly, Bob and Charlie obtain an
approximation of $\left| \varphi ^{/}\right\rangle $ with the maximal
fidelity $\frac{5}{6}$, respectively. In this case, the relative entropy of
entanglement in the cut of $a:B$ (or $a:C$) of the entangled state $\left|
\psi \right\rangle $, $E_{r}\approx 0.4425$.

We examine further in the RSP of a polar state the relation between the
entanglement and the fidelity under 1 bit classical communication, and
rewrite the Equations (6) and (7) in general form: 
\begin{equation}
\left| \Phi _{0}\right\rangle _{ABC}=\alpha \left| 0\right\rangle _{A}\left|
0\right\rangle _{B}\left| 0\right\rangle _{C}+\beta \left| 1\right\rangle
_{A}(\left| 0\right\rangle _{B}\left| 1\right\rangle _{C}+\left|
1\right\rangle _{B}\left| 0\right\rangle _{C})
\end{equation}
\begin{equation}
\left| \Phi _{1}\right\rangle _{ABC}=\alpha \left| 1\right\rangle _{A}\left|
1\right\rangle _{B}\left| 1\right\rangle _{C}+\beta \left| 0\right\rangle
_{A}(\left| 0\right\rangle _{B}\left| 1\right\rangle _{C}+\left|
1\right\rangle _{B}\left| 0\right\rangle _{C}),
\end{equation}
where $\alpha ,\beta $ are real numbers, and satisfy $\alpha ^{2}+2\beta
^{2}=1$. It can be deduced from the isotropy of the procedure that the
fidelity $F=\frac{1+\alpha ^{2}}{2}$,and $F_{\max }=\frac{5}{6}$. A simple
calculation gives 
\begin{equation}
E_{r}=\frac{3F-1}{2}\log (\frac{3F-1}{2})+\frac{1-F}{2}\log (\frac{1-F}{2}%
)-F\log (\frac{F}{2})
\end{equation}
The analytic relation can be plotted in Fig.1. It is obvious that the
required amount of entanglement increases with the fidelity in the limit of
certain classical communication cost. And it is noticed that the trade-off
relation ties in nicely with ideas concerning the relation between the
amount of classical information on a quantum state and the available
entanglement \cite{prl84}.

We consider another situation of preparing remotely two instances of quantum
state at the same location, where Alice holds only qubit $a$ and the qubits
A, B and C belong to Bob. It requires 1 bit classical communication from
Alice to Bob and 1 ebit in the cut of $a:ABC$. Instead, the task can be
achieved by the combination of the RSP of one instance of quantum state and
the optimal cloning procedure. In the two schemes the required amount of
entanglement is equal for the same optimal fidelity and the same 1 bit
classical communication.

In addition, there is something interesting in the scheme if we continue to
inspect the scheme in view of the RSP of a tripartite state. It is noticed
that after the RSP, a one-parameter tripartite entangled state $\left| \xi
\right\rangle _{ABC}$ is constructed among the three parties Alice, Bob and
Charlie. The parameter $\phi \in (0,2\pi ]$ (or $\theta \in (0,\pi ]$) is
dependent on Alice's choice. So the scheme can also be used to prepare
remotely a one-parameter tripartite state by Alice broadcasting 1 bit to and
pre-sharing entanglement with Bob and Charlie.

\section{Conclusion}

In this short note, we give a scheme in which two instances of an equatorial
state (or a polar state) can be remotely prepared to Bob and Charlie by
Alice broadcasting 1 bit to Bob and Charlie through public channel and
sharing prior entanglement among three parties. The trade-off curve between
the amount of entanglement and the achievable fidelity is derived under the
conditions of certain classical communication and constrained state. And in
view of the RSP of the tripartite state, the scheme can be used to prepare
remotely a one-parameter tripartite entangled state by prior entanglement
and 1 bit of broadcasting. It gives an example for the statement about the
preparation of the tripartite state in \cite{bennett}. The scheme can be
easily generalized to the case of multiple (more than two) receivers by
employing more parties entangled state. More generally, it inspires us to
study quantum remote control on multiple separate objects, for example,
teleportation of unitary operations\cite{pra63,pra65} to several different
receivers, and gives us an insight into the trade-offs between the
communication requirement and the restrictions about operations and quantum
states. We hope that it will shed some light on the understanding of the
fundamental laws of quantum information processing and the research of
quantum communication complexity.

\begin{center}
{\bf Acknowledgments}
\end{center}

This work has been financially supported by the National Natural Science
Foundation of China under the Grant No.10074072.

\end{document}